\documentclass{article}
\usepackage{spconf,amsmath,graphicx}
\usepackage{booktabs} % Table formatting
\usepackage{siunitx} % SI units

\title{Multi-task Learning for Radar Signal Characterisation}
\name{Zi Huang$^{\star \dagger}$, Akila Pemasiri$^{\star}$, Simon Denman$^{\star}$, Clinton Fookes$^{\star}$, Terrence Martin$^{\dagger}$}
\address{$^{\star}$Queensland University of Technology, Australia \\ $^{\dagger}$Revolution Aerospace, Australia}
\begin{document}
%\ninept
%
\maketitle
\begin{abstract}
Radio signal recognition is a crucial task in both civilian and military applications, as accurate and timely identification of unknown signals is an essential part of spectrum management and electronic warfare. The majority of research in this field has focused on applying deep learning for modulation classification, leaving the task of signal characterisation as an understudied area. This paper addresses this gap by presenting an approach for tackling radar signal classification and characterisation as a multi-task learning (MTL) problem. We propose the IQ Signal Transformer (IQST) among several reference architectures that allow for simultaneous optimisation of multiple regression and classification tasks. We demonstrate the performance of our proposed MTL model on a synthetic radar dataset, while also providing a first-of-its-kind benchmark for radar signal characterisation.

\end{abstract}
%

% -------------------------------------------------------------------------

% \vspace{-5pt}
\section{Introduction}
\label{sec:intro}
\vspace{-5pt}
Recent innovations in deep learning (DL) coupled with the declining cost of computation have enabled the successful application of deep neural networks (DNNs) for radio signal recognition (RSR). RSR can be defined as the process of extracting the hidden characteristics within the radio frequency (RF) waveform to aid in the identification of unknown radio emitters. This capability is foundational to both civilian and military integrated sensing and communications 
 (ISAC) applications \cite{haigh_cognitive_2021-1}, such as to improve the spectrum utilisation in communications networks and to enhance the spectrum situational awareness of soldiers in the modern battlefield.

Traditionally, classification of RF waveforms is achieved using likelihood-based \cite{wen_wei_maximum-likelihood_2000} and feature-based \cite{desimio_adaptive_1988, hsiao-chun_wu_novel_2008} methods that exploit the unique characteristics observed in the captured signal, such as its cyclostationary behaviour \cite{kim2007cyclostationary} and statistical features \cite{hsiao-chun_wu_novel_2008}. However, traditional methods are generally labour intensive requiring expert feature engineering and \textit{a priori} knowledge about the signal characteristics, and thus cannot effectively cater for non-cooperative and covert spectrum users \cite{haigh_cognitive_2021-1}. DL-based RSR solutions have garnered significant attention in recent years \cite{logue2019expert,vila2019deep} as they hold promise in effectively addressing these challenges. 

The application of convolutional neural networks (CNNs) to automatic modulation classification (AMC) was introduced by \cite{oshea_convolutional_2016-1}. Their early works \cite{oshea_unsupervised_2016-1, west_deep_2017-1} together with the release of several public datasets \cite{oshea_over--air_2018} initiated a wave of interest in DL-based RSR. Recently, several alternative DL approaches that adopt recurrent neural networks (RNNs) and hybrid architectures \cite{huynh-the_automatic_2021} were able to consistently achieve above 90\% modulation classification accuracy in relatively high signal-to-noise ratio (SNR) settings. Despite the success of DNNs, many recent approaches still rely on handcrafted features to pre-process the complex-valued, in-phase and quadrature (IQ) data into image-based representations, such as spectrograms \cite{huynh-the_automatic_2021}, prior to training. These approaches effectively transform RSR into an image classification problem, and thus limits the ability of DNNs to extract the fine-grained temporal relationships within the IQ data. While transformers \cite{vaswani2017attention} have found success in adjacent fields such as audio signal processing \cite{gong_ast_2021-1}, CNN-based models still dominate the current solution landscape for AMC.

Despite the progress made in RSR, the majority of recent research has only focused on AMC and wireless communication waveforms in a civilian context. While classifying modulation schemes can provide useful insight on the radio spectrum use, this information alone is insufficient in identifying or intercepting radio emitters, which is a highly desirable capability in a military context \cite{haigh_cognitive_2021-1}. Signal characterisation extends the scope of AMC by extracting additional signal characteristics, such as estimating the pulse width (PW) and pulse repetition interval (PRI) of a radar transmission. Specifically for radar signal characterisation (RSC), estimating the pulse descriptor words (PDWs) of radar systems is an essential part of electronic warfare. PDWs, which comprise specific radar signal parameters such as PW and PRI, are essential for constructing threat libraries. 

It is possible that limited DL research covering RSC may be attributed to the lack of publicly available radar datasets, as the majority of existing work in RSR has only focused on a single task, such as AMC \cite{huynh-the_automatic_2021}. Recently, multi-task learning (MTL) approaches to RSR were investigated in \cite{jagannath_multi-task_2021-1,jagannath_multi-task_2022-1}. These were the initial works that explored RSR as a joint problem by simultaneously classifying modulation and signal types on a synthetic radar and communication dataset \cite{jagannath_multi-task_2021-1}. While MTL was demonstrated in \cite{jagannath_multi-task_2022-1} to be effective at performing RSR in resource-constrained environments, this work was limited to classification tasks only. Furthermore, the proposed dataset lacks labelled signal characteristics that are required to support DL model development for RSC. 

To address the existing gaps, this paper introduces a MTL framework for RSC. In addition, we introduce the IQ Signal Transformer (IQST) to perform automatic feature extraction on IQ data without requiring handcrafted features or image-based transforms. Our main contributions are threefold. First, we produce a synthetic radar signals dataset with multiple categorical and numerical labels needed to support MTL. Our dataset will be made available for public use\footnote{The download link to our synthetic dataset can be accessed via GitHub at: https://github.com/abcxyzi/RadChar}. Second, we propose a novel MTL architecture for RSC solving classification and regression tasks as a joint problem. Finally, we introduce a new benchmark for RSC and provide several reference architectures for MTL.

% -------------------------------------------------------------------------

% \vspace{-5pt}
\section{Proposed Method}
\label{sec:method}
\vspace{-5pt}

% \subsection{Problem Definition}
% \label{ssec:problem_def}

\subsection{Dataset Generation}
\label{ssec:signalmodel}
\begin{figure}[tb]

\begin{minipage}[b]{1.0\linewidth}
  \centering
  \centerline{\includegraphics[width=8.5cm]{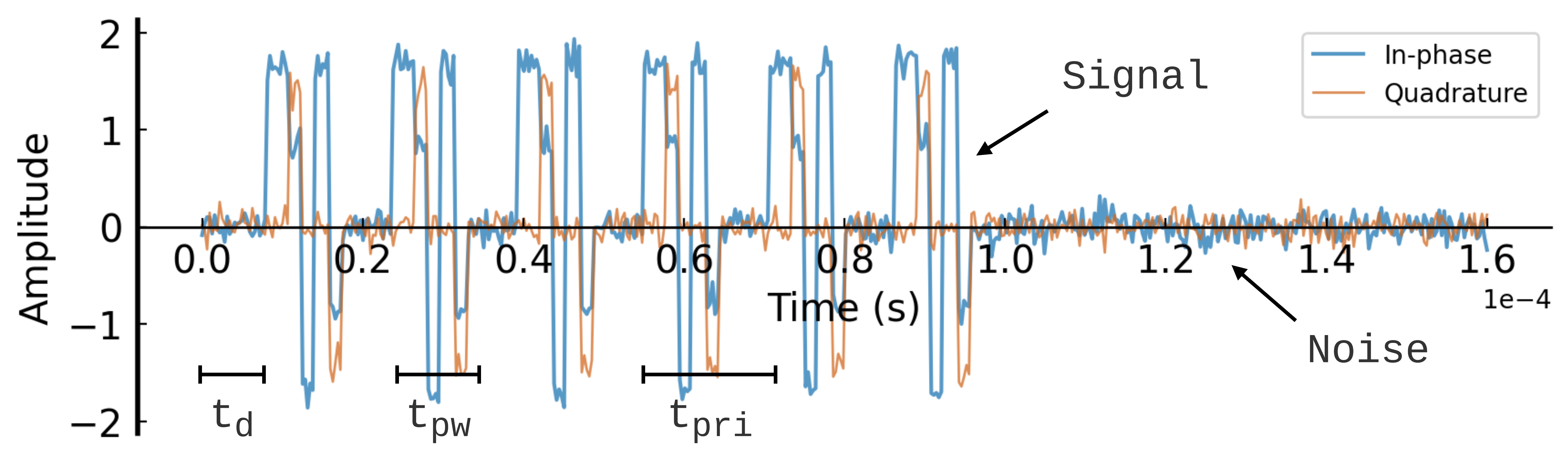}}
%  \vspace{2.0cm}
  \centerline{(a) SNR of 20 dB}\medskip
\end{minipage}
\begin{minipage}[b]{.48\linewidth}
  \centering
  \centerline{\includegraphics[width=4.0cm]{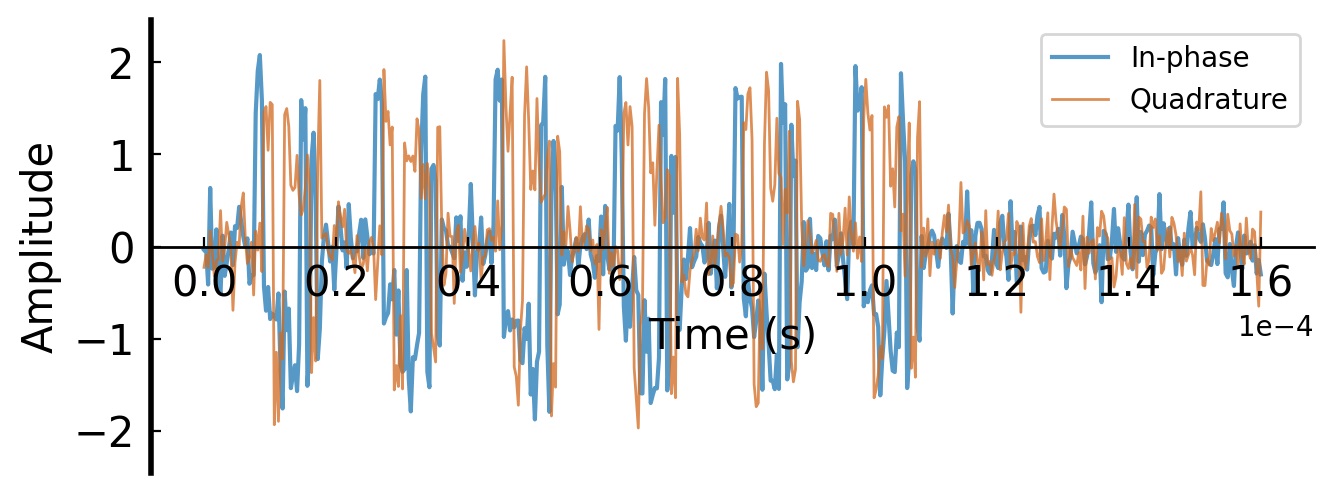}}
%  \vspace{1.5cm}
  \centerline{(b) SNR of 12 dB}\medskip
\end{minipage}
\hfill
\begin{minipage}[b]{0.48\linewidth}
  \centering
  \centerline{\includegraphics[width=4.0cm]{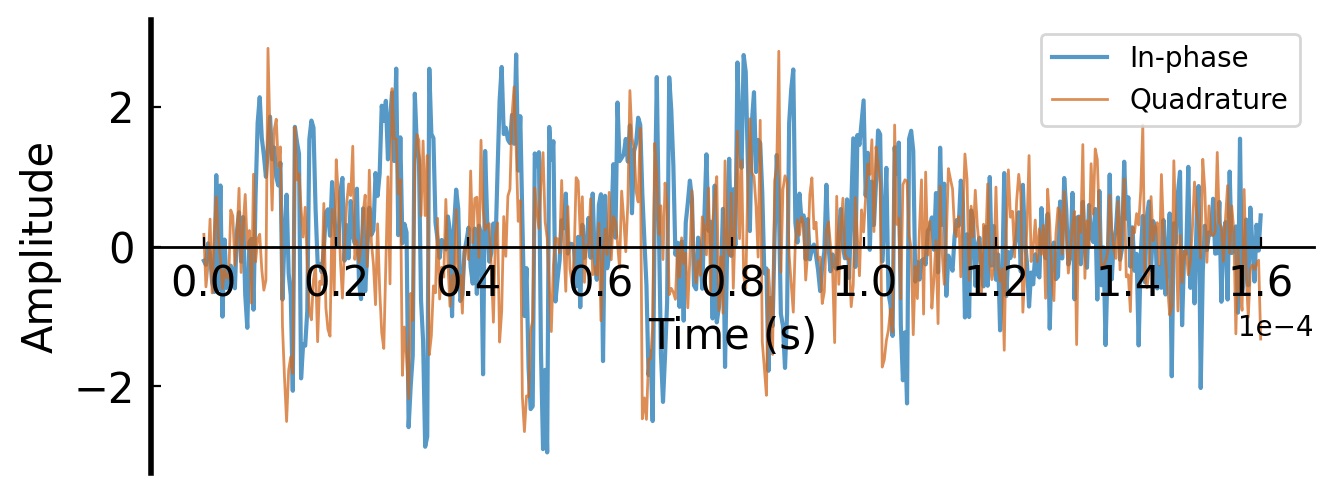}}
%  \vspace{1.5cm}
  \centerline{(c) SNR of 4 dB}\medskip
\end{minipage}
\caption{Radar signals sampled from the RadChar dataset illustrating polyphase Barker codes at varying SNRs.}
\label{fig:radar_waveform}
\end{figure}

\vspace{-5pt}
% We present a novel approach of using multi-task learning for radar signal characterisation. As this is the first study of its kind, we have generated synthetic radar data with suitable signal characteristics to support this study. 
Although existing datasets such as RadioML \cite{oshea_over--air_2018} and RadarComms \cite{jagannath_multi-task_2021-1} are useful for AMC, they do not provide training labels that are needed for RSC, and thus a new dataset is required. We generate our radar signals following the derivations in \cite{levanon2004radar}, at varying SNRs between -20 to 20 dB. To limit the number of signal parameters for the RSC problem, our dataset (RadChar) specifically focuses on pulsed radar signals. RadChar comprises a total of 5 radar signal types each covering 4 unique signal parameters. The signal classes include: Barker codes, polyphase Barker codes, Frank codes, linear frequency-modulated (LFM) pulses, and coherent unmodulated pulse trains. The signal parameters include PW ($t_{\text{pw}}$), PRI ($t_{\text{pri}}$), number of pulses ($n_{\text{p}}$), and pulse time delay ($t_{\text{d}}$). For phase-coded pulses, code lengths ($l_c$) of up to 13 and 16 are considered in Barker and Frank codes respectively. A radar waveform example is shown in Figure \ref{fig:radar_waveform}.

We carefully design each waveform in RadChar to contain 512 baseband IQ samples ($\Vec{x}_{\text{i}} + j \Vec{x}_{\text{q}}$) while ensuring the range of radar parameter values used to construct the dataset adheres to the Nyquist-Shannon sampling theorem. The minimum sampling frequency ($f_{\text{s}}$) required as a function of the selected radar characteristics is given by (\ref{eq:sampling_rate}). The sampling rate used in RadChar is 3.2 MHz. The numerical bounds selected for radar parameters $t_{\text{pw}}$, $t_{\text{pri}}$, $t_{\text{d}}$ and $n_{\text{p}}$ are 10-16 $\si{\micro\second}$, 17-23 $\si{\micro\second}$, 1-10 $\si{\micro\second}$, and 2-6 respectively. We apply uniformly random sampling across these value ranges for each signal class to generate 1 million unique radar waveforms. Additive white Gaussian noise (AWGN) is used to simulate varying SNRs in the dataset. In addition, we impose a unity average power to each radar waveform to ensure that signal power is scaled consistently across the dataset.

\begin{equation}
\label{eq:sampling_rate}
    % sps > 2 \times max\left ( \frac{l_{\text{c}}}{t_{\text{pw}}}, \frac{1}{\delta t}, \frac{1}{t_{\text{pri}}}\right )
    % sps > 2 \times max \left \{ l_{\text{c}} t_{\text{pw}}^{-1}, t_{\text{pri}}^{-1}, t_{\text{d}}^{-1} \right \}
    % f_{\text{s}} > 2 \cdot \max ( l_{\text{c}} t_{\text{pw}}^{-1}, t_{\text{pri}}^{-1}, t_{\text{d}}^{-1} )
    f_{\text{s}} > 2 \cdot \max ( l_{\text{c}} t_{\text{pw}}^{-1}, t_{\text{pri}}^{-1}, t_{\text{d}}^{-1} )
    % f_{\text{s}} > 2B \rightarrow f_{\text{s}} > 2 (max ( l_{\text{c}} t_{\text{pw}}^{-1}, t_{\text{pri}}^{-1}, t_{\text{d}}^{-1} ))
\end{equation}

\vspace{-5pt}
\subsection{Multi-task Learning Framework}
\label{ssec:mtl}
\vspace{-5pt}
The proposed MTL model for RSC adopts the hard parameter shared MTL approach \cite{caruana_multitask_1993-1}, where individual tasks share a single neural network backbone. Our model comprises two segments: a modular backbone for learning shared representations on the raw IQ data, and a set of parallel task-specific heads which consist of classification and regression tasks for signal classification and characterisation respectively. Our approach benefits from its modularity as the choice of the shared backbone is flexible allowing for domain adaptation, while additional task-specific heads can be added to increase the scope of the model. Furthermore, hard parameter sharing is advantageous for learning common representations of similar tasks, such as RSC tasks, and significantly reduces the risk of overfitting as the number of related tasks increases \cite{baxter_bayesianinformation_1997}. 

For the multi-task segment of our model, we follow the approach in \cite{jagannath_multi-task_2021-1} to construct task-specific heads using a minimal set of hidden layers. To achieve a lightweight design, all task-specific heads are the same depth and each contains a single convolutional layer with a kernal size of 3$\times$3 followed by a dense layer. Dropout rates of 0.25 and 0.5 are applied to convolutional and dense layers respectively. The number of convolutional filters used here is driven by the output dimension of the shared backbone. We adopt the ReLU activation function in each head, while batch normalisation is applied prior to the activation function. Our model contains 5 task-specific heads which include a single classification head for signal classification, and 4 regression heads for signal characterisation. For classification, a softmax function is used to output probabilities for individual signal classes, while parameter predictions are obtained directly from the dense layer for each regression task. 

The proposed MTL model is trained by optimising a compound multi-task loss ($L_{\text{mtl}}$) function given by (\ref{eq:mtl_loss}). % as a weighted sum of losses over individual tasks ($i$). 
The classification task is optimised using a categorical cross-entropy loss function, while the regression tasks are optimised using an L1 loss function. The multi-task loss is parameterised by shared parameters ($\theta_{\text{sh}}$) from the model backbone and task-specific parameters ($\theta_{1},...,\theta_{5}$) from individual task heads. The weights ($w_i$) of task-specific losses are MTL hyperparameters and joint optimisation, given by (\ref{eq:opt}), is achieved by minimising the total loss from task-specific heads. %Here, we consider signal type estimation as a classification task and signal parameter estimation as regression tasks. 

\begin{equation}
\label{eq:mtl_loss}
    L_{\text{mtl}}(\theta_{\text{sh}}, \theta_{1}, ...,  \theta_{5}) = \sum_{i=1}^{5} w_{i}L_{i}(\theta_{\text{sh}}, \theta_{i})
\end{equation}

\begin{equation}
\label{eq:opt}
    \operatorname*{argmin}_{\theta_{\text{sh}}, \theta_{1}, ..., \theta_{5}} L_{\text{mtl}}(\theta_{\text{sh}}, \theta_{1}, ...,  \theta_{5})
\end{equation}

\begin{figure}[tb]
    \centering
    \includegraphics[width=0.44\textwidth]{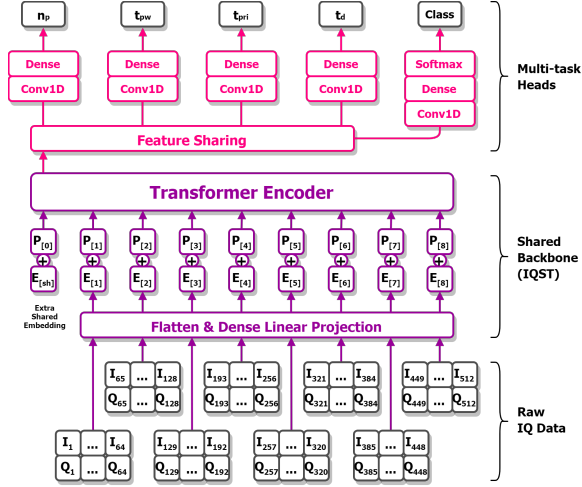}
    \caption{The proposed hard parameter shared MTL architecture for RSC. This model shows an IQST backbone with task-specific classification and regression heads.}
\label{fig:mtl_net}
\end{figure}

\vspace{-5pt}
\subsection{Shared Feature Extraction Backbones}
\label{ssec:model_archs}
\vspace{-5pt}
We provide several reference designs of the MTL backbone to perform feature extraction on the raw IQ data. The uniqueness of our approach is that our models operate directly on raw IQ data requiring no additional pre-processing and feature transforms as seen in \cite{logue2019expert,vila2019deep}. Because our MTL architecture is inherently modular, our models can easily be extended to incorporate additional classification and regression tasks in order to increase the scope of RSC. 

We provide two CNN implementations of the shared feature extraction backbone. CNN2D follows the same design philosophy as \cite{jagannath_multi-task_2021-1} to achieve a lightweight model. It comprises a single convolution layer with 8 filters using a kernal size of 2$\times$2 followed by a 2$\times$2 max pooling operation. Input to the CNN is a single channel 32$\times$32 tensor which is reshaped from the raw IQ data. By intuition, such an uninformed reshaping operation is non-ideal for representing IQ data that is inherently sequential. We propose a modification to this approach by directly ingesting the IQ data as two separate I and Q channels of shape 2$\times$512 to retain the shape of the raw IQ sequence. CNN1D uses 1D convolutional and max pooling operations instead, while maintaining the same number of filters as CNN2D. ReLU is used as the activation function in both backbones with a dropout rate of 0.25. 

We introduce the IQ Signal Transformer (IQST), as shown in Figure \ref{fig:mtl_net}, a novel attention-based architecture tailored for RSC and MTL. Our design is inspired by the Audio Spectrogram Transformer (AST) from \cite{gong_ast_2021-1}, though adopts the standard transformer encoder architecture from \cite{vaswani2017attention}. Unlike \cite{gong_ast_2021-1}, our approach operates on the raw signal allowing for direct feature extraction from the IQ data without the need to first transform the IQ data to an image representation. We adopt the patch embedding technique of \cite{gong_ast_2021-1,dosovitskiy_image_2021} to generate a sequence of 1D patch embeddings from a 2$\times$512 tensor constructed from the raw IQ sequence. The dual-channel IQ data is flattened to form 8$\times$1$\times$128 blocks (or tokens) prior to applying a dense linear projection to form 8 learnable patch embeddings, each with an embedding dimension of 768. Each embedded patch is added to the standard positional embeddings from \cite{vaswani2017attention} to form a 128$\times$8 input to the transformer encoder. We include an additional learnable embedding to the encoder to allow for common feature sharing across the individual tasks. This extra embedding is similar to the class embedding from \cite{dosovitskiy_image_2021}. The standard IQST (IQST-S) adopts the GELU activation function and implements 3 multi-head attention blocks and 3 encoder layers. We feed the outputs from the shared embedding as a 1$\times$128 feature map into each task-specific head to complete the MTL model. 

\begin{table*}[tb]
\noindent\begin{tabular*}{\textwidth}
{@{\hspace{0.2cm}}@{\extracolsep{\stretch{1}}}*{6}{r}@{\hspace{0.2cm}}}
    \toprule
    \multicolumn{1}{c}{Model} & \multicolumn{1}{c}{MAE($n_{\text{p}}$)} & \multicolumn{1}{c}{MAE($t_{\text{pw}}$)} & \multicolumn{1}{c}{MAE($t_{\text{pri}}$)} & \multicolumn{1}{c}{MAE($t_{\text{d}}$)} & \multicolumn{1}{c}{Class Acc.} \\
    % Model & MAE($n_{\text{p}}$) & MAE($t_{\text{pw}}$) & MAE($t_{\text{pri}}$) & MAE($t_{\text{d}}$) & Class Acc. \\
    \midrule
    \multicolumn{1}{c}{CNN1D} & \textbf{0.729}, 0.193, \textbf{0.085} & 1.413, \textbf{0.560}, 0.340 & 0.999, 0.330, 0.209 & 1.349, 0.385, \textbf{0.206} & 0.757, 0.998, \textbf{1.000} \\
    \multicolumn{1}{c}{CNN2D} & 0.793, \textbf{0.174}, 0.090 & 1.466, 0.801, 0.505 & 1.054, 0.420, 0.299 & 1.729, 0.638, 0.443 & 0.673, 0.983, 0.998 \\
    % LSTM & 0.00 & 0.39 & 0.02 & 0.49 \\
    \multicolumn{1}{c}{IQST-S} & 0.733, 0.294, 0.251 & 1.282, 0.628, 0.364 & 0.816, \textbf{0.273}, \textbf{0.192} & \textbf{1.229}, 0.415, 0.277 & \textbf{0.792}, \textbf{0.999}, \textbf{1.000} \\
    \multicolumn{1}{c}{IQST-L} & 0.752, 0.195, 0.124 & \textbf{1.253}, 0.579, \textbf{0.334} & \textbf{0.799}, 0.286, 0.225 & 1.253, \textbf{0.379}, 0.233 & 0.791, 0.998, \textbf{1.000} \\
    \bottomrule
\end{tabular*}
\caption{Comparison of task performance across MTL models. Each value string (-,-,-) shows the performance of each task at -10, 0 and 10 dB SNR respectively. A lower MAE value is desired for regression, while a higher accuracy value indicates better classification performance. Note that the units for $t_{\text{pw}}$, $t_{\text{pri}}$ and $t_{\text{d}}$ are expressed in $\si{\micro\second}$.}
\label{table:results}
\end{table*}

\begin{figure*}[tb]

\begin{minipage}[b]{0.19\linewidth}
  \centering
  \centerline{\includegraphics[width=3.2cm]{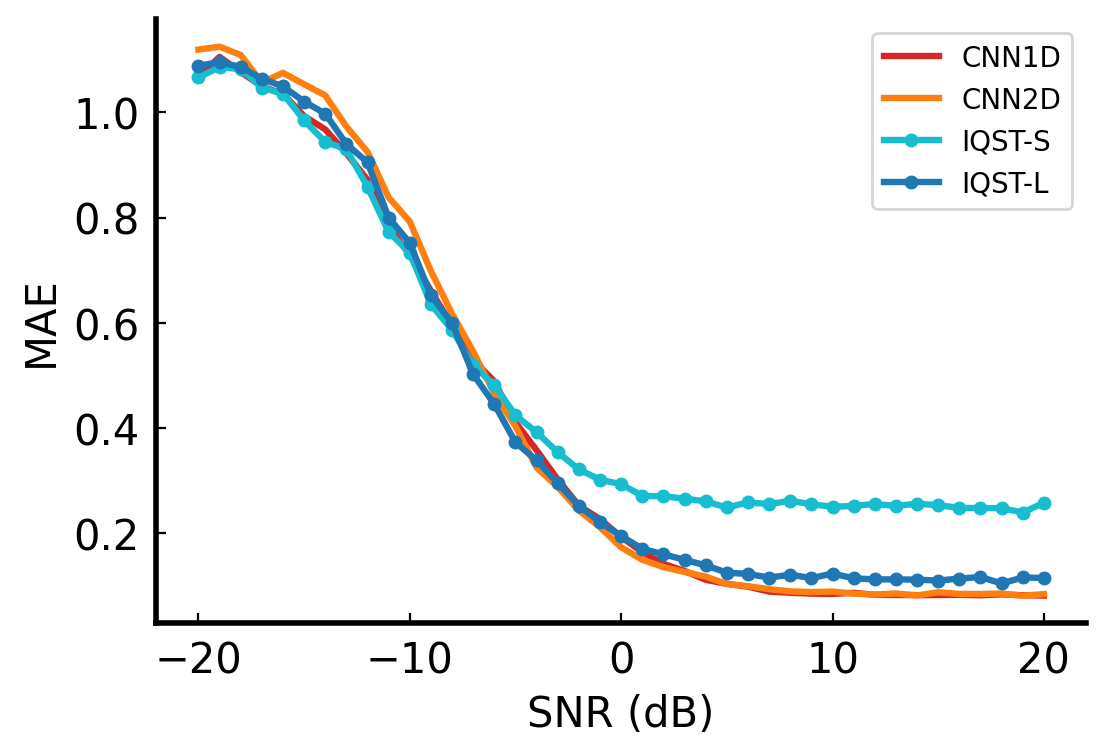}}
  \centerline{(a) MAE($n_{\text{p}}$)}\medskip
\end{minipage}
\hfill
\begin{minipage}[b]{0.19\linewidth}
  \centering
  \centerline{\includegraphics[width=3.2cm]{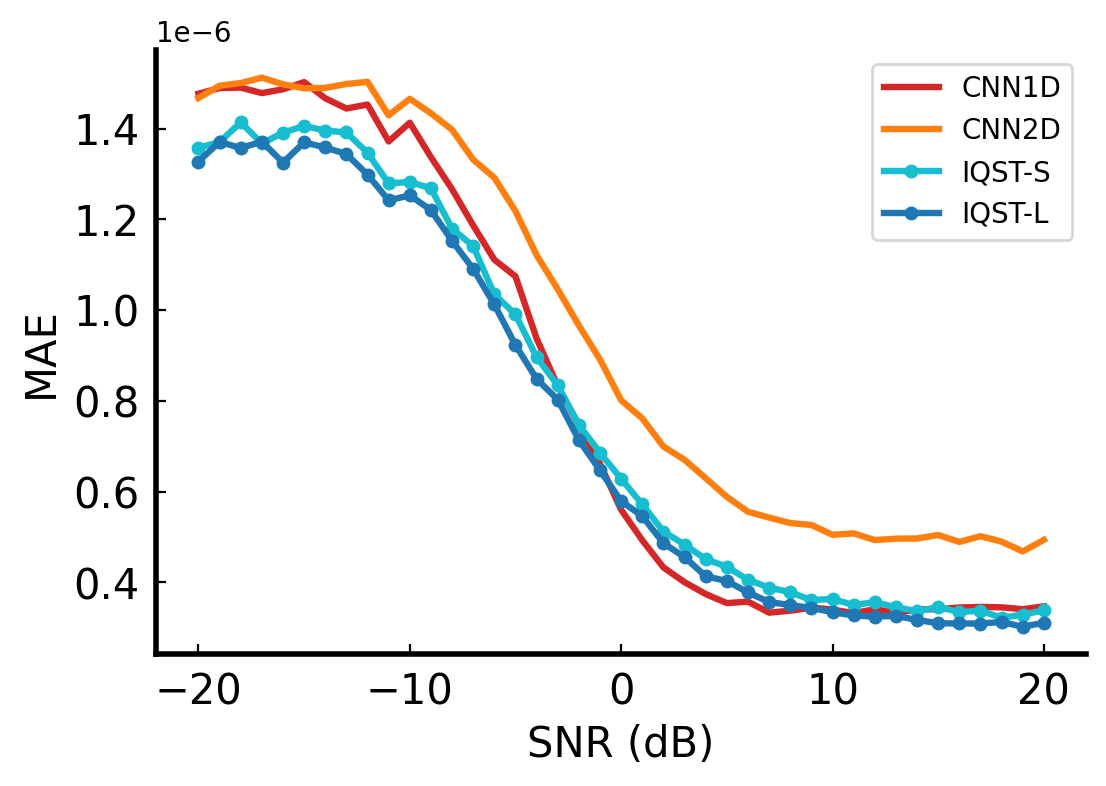}}
  \centerline{(b) MAE($t_{\text{pw}}$)}\medskip
\end{minipage}
\hfill
\begin{minipage}[b]{0.19\linewidth}
  \centering
  \centerline{\includegraphics[width=3.2cm]{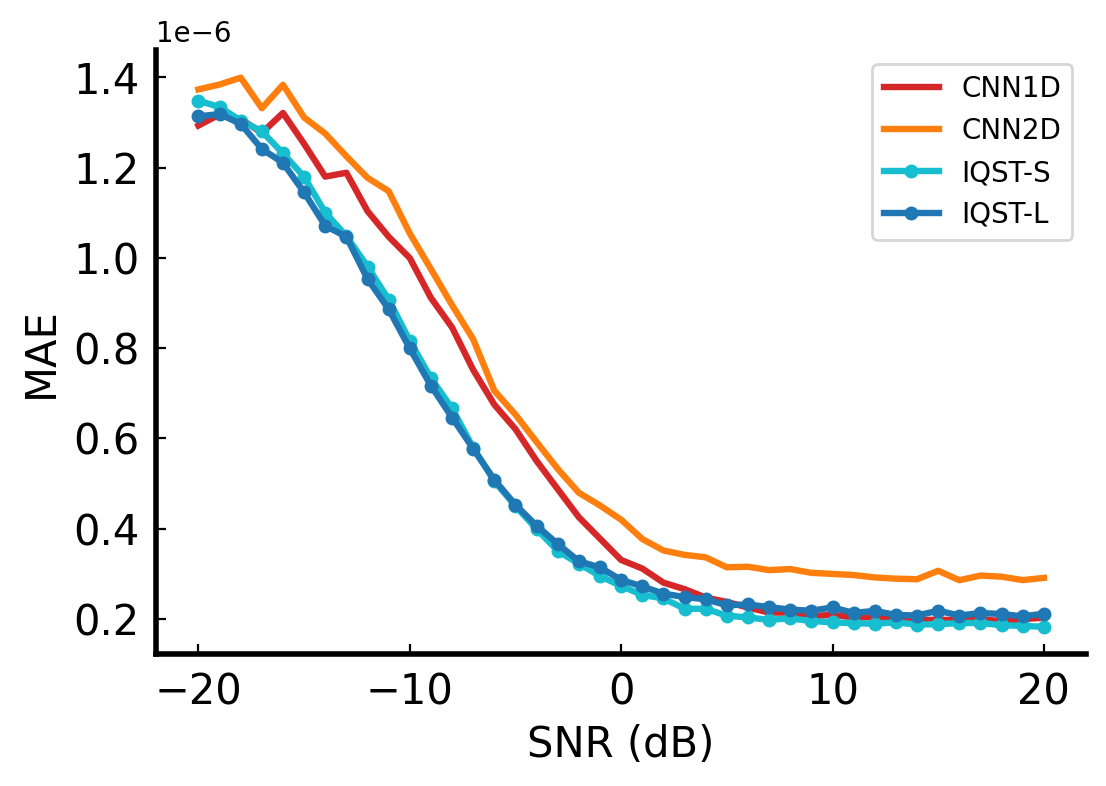}}
  \centerline{(c) MAE($t_{\text{pri}}$)}\medskip
\end{minipage}
\hfill
\begin{minipage}[b]{0.19\linewidth}
  \centering
  \centerline{\includegraphics[width=3.2cm]{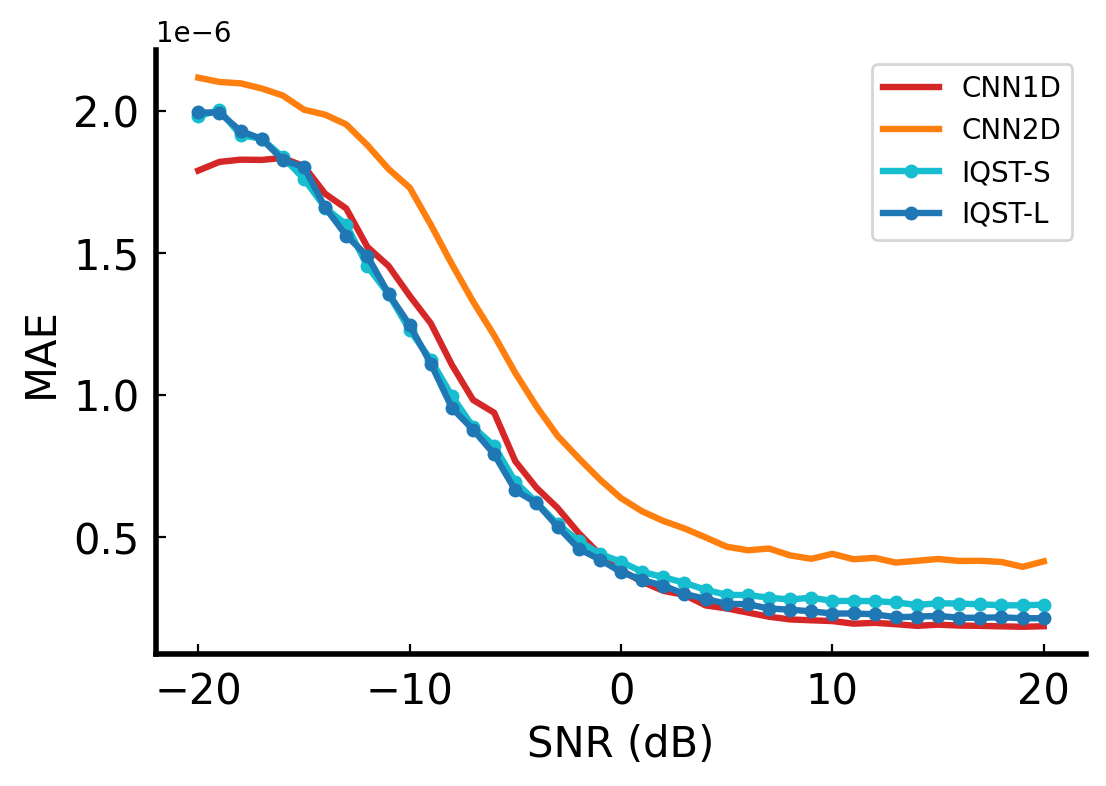}}
  \centerline{(c) MAE($t_{\text{d}}$)}\medskip
\end{minipage}
\hfill
\begin{minipage}[b]{0.19\linewidth}
  \centering
  \centerline{\includegraphics[width=3.2cm]{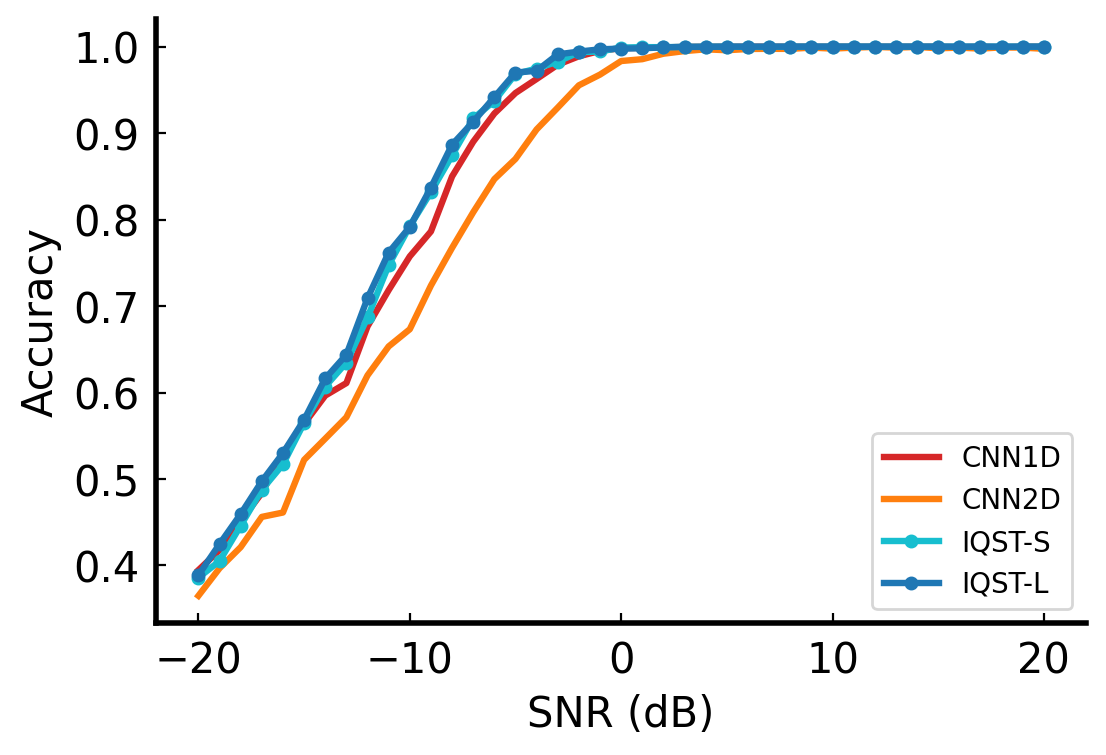}}
  \centerline{(e) Class Acc.}\medskip
\end{minipage}
\caption{Test performance of MTL models across an SNR range of -20 to 20 dB. MAE results of the regression tasks are shown in (a) to (d), and signal type classification accuracy is shown in (e).}
\label{fig:class_acc}
\end{figure*}

% \vspace{-5pt}
\section{Experiments}
\label{sec:results}
\vspace{-5pt}

\subsection{Training Details}
\label{ssec:exp_details}
\vspace{-5pt}
We train and evaluate our models on a single Nvidia Tesla A100 GPU. A 70-15-15\% train-validation-test split of RadChar is used for all our experiments. We train our models for 100 epochs with a learning rate of 5e-4 and a batch size of 64. We adopt the Adam optimiser and initialise the model parameters using LeCun initialisation. Importantly, we standardise the raw IQ samples against the training population mean and variance, and also normalise the regression labels between 0 and 1. The latter step significantly improves training performance and convergence for regression tasks, especially when dealing with small time values such as radar parameters.

\vspace{-5pt}
\subsection{MTL Model Performance}
\label{ssec:model_perf}
\vspace{-5pt}
We evaluate our MTL models on the RadChar dataset. Model performance is compared on the same test set using task-specific metrics. Classification accuracy and mean absolute error (MAE) are selected to evaluate the performance of classification and regression tasks respectively. Table \ref{table:results} provides a summary of individual task performance across various SNRs. The performance of a larger IQST (IQST-L), which uses 9 multi-head attention blocks and 6 encoder layers is also shown here for comparison. While CNN2D underperforms against other models across all tasks, IQST models generally perform better, especially at low SNRs. We observe from Figure \ref{fig:class_acc} that as SNR increases, MAE decreases and classification accuracy improves, with the latter trend consistent with what is expected for AMC \cite{oshea_convolutional_2016-1,west_deep_2017-1}. %This trend is highlighted in Figure \ref{fig:class_acc}.

The outstanding performance of 1D models substantiates the importance of representing IQ data as 1D sequences. IQST benefits from its transformer architecture, which better captures longer-term dependencies between IQ samples, therefore allowing it to perform better at low SNRs. Although CNN1D outperforms IQST-S in the $n_{\text{p}}$ estimation task at high SNRs, increasing the model capacity, as in IQST-L, shows a significant improvement in task performance. While a larger transformer encoder is capable of capturing more complex dependencies in the IQ sequence, the computational cost is significantly increased due to its quadratic bottleneck \cite{vaswani2017attention}. Additionally, our results highlight the challenge in MTL, where a trade-off in task performance may need to be considered as the number of individual tasks increases. Nevertheless, our results indicate the potential for attention-based, hard parameter-shared MTL models for RSC.

\vspace{-5pt}
\subsection{Ablation Study}
\label{ssec:ablation}
\vspace{-5pt}
Increasing the number of convolutional layers did not provide a notable improvement on individual task performance. Instead, we find that deeper convolutional networks negatively impact regression tasks and result in higher errors. % We hypothesise that the resolution of the IQ data is not nearly as complex when compared to typical computer vision problems in order to benefit from deeper networks.
We hypothesise that regression tasks which require accurate estimation of time parameters are adversely affected by the stacking of operations, which reduces temporal resolution. % The reduced temporal resolution of IQST resulting from the tokenisation of the raw signal, perhaps a sliding window would help here. Transformers lack the inductive bias inherent to CNNs and therefore do not generalise well when trained on small datasets, perhaps more data would help.
% We apply uniformly random sampling to select task weights to study their impact on $L_{\text{mt}}$. 
Separately, selecting task weights that produce a relatively even distribution of $w_iL_i$ during model initialisation provides stable task performance over all SNRs, while increasing the task weight to favour a specific task did not appear to help improve its test performance. This is true for both classification and regression tasks. Our observations are consistent with similar findings from \cite{jagannath_multi-task_2021-1} under the same SNR environment. A weight distribution of 0.1 for classification and 0.225 for all regression tasks was used in the experiments shown. 

% -------------------------------------------------------------------------

\vspace{-5pt}
\section{Conclusion}
\label{sec:conclusion}
\vspace{-5pt}
In this paper, we present a MTL framework for tackling RSC as a joint optimisation problem. We propose the IQST among other reference architectures to perform simultaneous optimisation of classification and regression tasks while highlighting the benefits of IQST for feature extraction on raw IQ data, particularly at low SNRs. We demonstrate the performance of our models on a synthetic radar dataset and provide a first-of-its-kind benchmark for RSC. The modularity of our proposed MTL design provides opportunities for additional classification and regression tasks in future work.

% -------------------------------------------------------------------------

% \vspace{-5pt}
\section{Acknowledgement}
\label{sec:acknowledgement}
\vspace{-5pt}
The research for this paper received funding support from the Queensland Government through Trusted Autonomous Systems (TAS), a Defence Cooperative Research Centre funded through the Commonwealth Next Generation Technologies Fund and the Queensland Government.

\vfill
\pagebreak

% -------------------------------------------------------------------------

\bibliographystyle{IEEEbib}
\bibliography{refs_cameraready}

\end{document}